
\font\twelverm=cmr10  scaled 1200   \font\twelvei=cmmi10  scaled 1200
\font\twelvesy=cmsy10 scaled 1200   \font\twelveex=cmex10 scaled 1200
\font\twelvebf=cmbx10 scaled 1200   \font\twelvesl=cmsl10 scaled 1200
\font\twelvett=cmtt10 scaled 1200   \font\twelveit=cmti10 scaled 1200
\font\twelvesc=cmcsc10 scaled 1200  \font\twelvesf=cmss10 scaled 1200
\skewchar\twelvei='177   \skewchar\twelvesy='60


\def\twelvepoint{\normalbaselineskip=12.4pt plus 0.1pt minus 0.1pt
  \abovedisplayskip 12.4pt plus 3pt minus 9pt
  \belowdisplayskip 12.4pt plus 3pt minus 9pt
  \abovedisplayshortskip 0pt plus 3pt
  \belowdisplayshortskip 7.2pt plus 3pt minus 4pt
  \smallskipamount=3.6pt plus1.2pt minus1.2pt
  \medskipamount=7.2pt plus2.4pt minus2.4pt
  \bigskipamount=14.4pt plus4.8pt minus4.8pt
  \def\rm{\fam0\twelverm}          \def\it{\fam\itfam\twelveit}%
  \def\sl{\fam\slfam\twelvesl}     \def\bf{\fam\bffam\twelvebf}%
  \def\mit{\fam 1}                 \def\cal{\fam 2}%
  \def\sc{\twelvesc}               \def\tt{\twelvett}
  \def\sf{\twelvesf}
  \textfont0=\twelverm   \scriptfont0=\tenrm   \scriptscriptfont0=\sevenrm
  \textfont1=\twelvei    \scriptfont1=\teni    \scriptscriptfont1=\seveni
  \textfont2=\twelvesy   \scriptfont2=\tensy   \scriptscriptfont2=\sevensy
  \textfont3=\twelveex   \scriptfont3=\twelveex  \scriptscriptfont3=\twelveex
  \textfont\itfam=\twelveit
  \textfont\slfam=\twelvesl
  \textfont\bffam=\twelvebf \scriptfont\bffam=\tenbf
  \scriptscriptfont\bffam=\sevenbf
  \normalbaselines\rm}



\def\beginlinemode{\endmode
  \begingroup\parskip=0pt \obeylines\def\\{\par}\def\endmode{\par\endgroup}}
\def\beginparmode{\endmode
  \begingroup \def\endmode{\par\endgroup}}
\let\endmode=\par
{\obeylines\gdef\
{}}
\def\singlespace{\baselineskip=\normalbaselineskip}

\def\oneandahalfspace{\baselineskip=\normalbaselineskip
  \multiply\baselineskip by 3 \divide\baselineskip by 2}
\def\doublespace{\baselineskip=\normalbaselineskip \multiply\baselineskip by 2}

\newcount\firstpageno
\firstpageno=2
\footline={\ifnum\pageno<\firstpageno{\hfil}\else{\hfil\twelverm\folio\hfil}\fi}
\def\toppageno{\global\footline={\hfil}\global\headline
  ={\ifnum\pageno<\firstpageno{\hfil}\else{\hfil\twelverm\folio\hfil}\fi}}
\let\rawfootnote=\footnote              
\def\footnote#1#2{{\rm\singlespace\parindent=0pt\parskip=0pt
  \rawfootnote{#1}{#2\hfill\vrule height 0pt depth 6pt width 0pt}}}
\def\raggedcenter{\leftskip=4em plus 12em \rightskip=\leftskip
  \parindent=0pt \parfillskip=0pt \spaceskip=.3333em \xspaceskip=.5em
  \pretolerance=9999 \tolerance=9999
  \hyphenpenalty=9999 \exhyphenpenalty=9999 }
\def\dateline{\rightline{\ifcase\month\or
  January\or February\or March\or April\or May\or June\or
  July\or August\or September\or October\or November\or December\fi
  \space\number\year}}
\def\received{\vskip 3pt plus 0.2fill
 \centerline{\sl (Received\space\ifcase\month\or
  January\or February\or March\or April\or May\or June\or
  July\or August\or September\or October\or November\or December\fi
  \qquad, \number\year)}}


\hsize=6.5truein
\hoffset=0.0truein
\vsize=8.9truein
\voffset=0.0truein
\parskip=\medskipamount
\def\\{\cr}
\twelvepoint            
\doublespace            
\overfullrule=0pt       


\newcount\timehour
\newcount\timeminute
\newcount\timehourminute
\def\daytime{\timehour=\time\divide\timehour by 60
  \timehourminute=\timehour\multiply\timehourminute by-60
  \timeminute=\time\advance\timeminute by \timehourminute
  \number\timehour:\ifnum\timeminute<10{0}\fi\number\timeminute}
\def\today{\number\day\space\ifcase\month\or Jan\or Feb\or Mar
  \or Apr\or May\or Jun\or Jul\or Aug\or Sep\or Oct\or
  Nov\or Dec\fi\space\number\year}




\def\ubec#1{
 \rightline{\rm UB--ECM--PF--#1}}      


\def\title                      
  {\null\vskip 3pt plus 0.2fill
   \beginlinemode \doublespace \raggedcenter \bf}

\def\author                     
  {\vskip 3pt plus 0.2fill \beginlinemode
   \doublespace \raggedcenter}

\def\affil                      
  {\vskip 3pt plus 0.1fill \beginlinemode
   \oneandahalfspace \raggedcenter \it}

\def\abstract                   
  {\vskip 3pt plus 0.3fill \beginparmode \narrower
   \oneandahalfspace {\it  Abstract}:\  }

\def\endtopmatter               
  {\endpage                     
   \body}

\def\body                       
  {\beginparmode}               

\def\head#1{                    
  \goodbreak\vskip 0.4truein    
  {\immediate\write16{#1}
   \raggedcenter {\sc #1} \par }
   \nobreak\vskip 0truein\nobreak}

\def\beneathrel#1\under#2{\mathrel{\mathop{#2}\limits_{#1}}}

\def\refto#1{$^{#1}$}           

\def\references                 
  {\head{References}            
   \beginparmode
   \frenchspacing \parindent=0pt    
   \parskip=0pt \everypar{\hangindent=20pt\hangafter=1}}

\gdef\refis#1{\item{#1.\ }}                     

\gdef\journal#1,#2,#3,#4.{              
    {\sl #1~}{\bf #2}, #3 (#4).}                

\def\endreferences{\body}

\def\figurecaptions             
  {\endpage
   \beginparmode
   \head{Figure Captions}
}

\def\endpage                    
  {\vfill\eject}

\def\endpaper                   
  {\endmode\vfill\supereject}


\def\ref#1{Ref.~#1}                     
\def\Ref#1{Ref.~#1}                     
\def\[#1]{[\cite{#1}]}
\def\cite#1{{#1}}
\def\(#1){(\call{#1})}
\def\call#1{{#1}}
\def\taghead#1{}
\def\frac#1#2{{#1 \over #2}}

\def\12{{1\over2}}

\def\sla{\raise.15ex\hbox{$/$}\kern-.57em}
\def\leaderfill{\leaders\hbox to 1em{\hss.\hss}\hfill}
\def\twiddle{\lower.9ex\rlap{$\kern-.1em\scriptstyle\sim$}}
\def\bigtwiddle{\lower1.ex\rlap{$\sim$}}
\def\gtwid{\mathrel{\raise.3ex\hbox{$>$\kern-.75em\lower1ex\hbox{$\sim$}}}}
\def\ltwid{\mathrel{\raise.3ex\hbox{$<$\kern-.75em\lower1ex\hbox{$\sim$}}}}
\def\square{\kern1pt\vbox{\hrule height 1.2pt\hbox{\vrule width 1.2pt\hskip 3pt
   \vbox{\vskip 6pt}\hskip 3pt\vrule width 0.6pt}\hrule height 0.6pt}\kern1pt}
\def\tdot#1{\mathord{\mathop{#1}\limits^{\kern2pt\ldots}}}

\def\pmb#1{\setbox0=\hbox{#1}%
  \kern-.025em\copy0\kern-\wd0
  \kern  .05em\copy0\kern-\wd0
  \kern-.025em\raise.0433em\box0 }


\def\crgrant{This research was supported in part by the CICYT
grant AEN90-0033. R.T. also acknowledges a grant from the
Ministerio de Educaci\'on y Ciencia, Spain}

\catcode`@=11
\newcount\tagnumber\tagnumber=0

\immediate\newwrite\eqnfile
\newif\if@qnfile\@qnfilefalse
\def\write@qn#1{}
\def\writenew@qn#1{}
\def\w@rnwrite#1{\write@qn{#1}\message{#1}}
\def\@rrwrite#1{\write@qn{#1}\errmessage{#1}}

\def\taghead#1{\gdef\t@ghead{#1}\global\tagnumber=0}
\def\t@ghead{}

\expandafter\def\csname @qnnum-3\endcsname
  {{\t@ghead\advance\tagnumber by -3\relax\number\tagnumber}}
\expandafter\def\csname @qnnum-2\endcsname
  {{\t@ghead\advance\tagnumber by -2\relax\number\tagnumber}}
\expandafter\def\csname @qnnum-1\endcsname
  {{\t@ghead\advance\tagnumber by -1\relax\number\tagnumber}}
\expandafter\def\csname @qnnum0\endcsname
  {\t@ghead\number\tagnumber}
\expandafter\def\csname @qnnum+1\endcsname
  {{\t@ghead\advance\tagnumber by 1\relax\number\tagnumber}}
\expandafter\def\csname @qnnum+2\endcsname
  {{\t@ghead\advance\tagnumber by 2\relax\number\tagnumber}}
\expandafter\def\csname @qnnum+3\endcsname
  {{\t@ghead\advance\tagnumber by 3\relax\number\tagnumber}}

\def\equationfile{%
  \@qnfiletrue\immediate\openout\eqnfile=\jobname.eqn%
  \def\write@qn##1{\if@qnfile\immediate\write\eqnfile{##1}\fi}
  \def\writenew@qn##1{\if@qnfile\immediate\write\eqnfile
    {\noexpand\tag{##1} = (\t@ghead\number\tagnumber)}\fi}
}

\def\callall#1{\xdef#1##1{#1{\noexpand\call{##1}}}}
\def\call#1{\each@rg\callr@nge{#1}}

\def\each@rg#1#2{{\let\thecsname=#1\expandafter\first@rg#2,\end,}}
\def\first@rg#1,{\thecsname{#1}\apply@rg}
\def\apply@rg#1,{\ifx\end#1\let\next=\relax%
\else,\thecsname{#1}\let\next=\apply@rg\fi\next}

\def\callr@nge#1{\calldor@nge#1-\end-}
\def\callr@ngeat#1\end-{#1}
\def\calldor@nge#1-#2-{\ifx\end#2\@qneatspace#1 %
  \else\calll@@p{#1}{#2}\callr@ngeat\fi}
\def\calll@@p#1#2{\ifnum#1>#2{\@rrwrite{Equation range #1-#2\space is bad.}
\errhelp{If you call a series of equations by the notation M-N, then M and
N must be integers, and N must be greater than or equal to M.}}\else%
 {\count0=#1\count1=#2\advance\count1
by1\relax\expandafter\@qncall\the\count0,%
  \loop\advance\count0 by1\relax%
    \ifnum\count0<\count1,\expandafter\@qncall\the\count0,%
  \repeat}\fi}

\def\@qneatspace#1#2 {\@qncall#1#2,}
\def\@qncall#1,{\ifunc@lled{#1}{\def\next{#1}\ifx\next\empty\else
  \w@rnwrite{Equation number \noexpand\(>>#1<<) has not been defined yet.}
  >>#1<<\fi}\else\csname @qnnum#1\endcsname\fi}

\let\eqnono=\eqno
\def\eqno(#1){\tag#1}
\def\tag#1$${\eqnono(\displayt@g#1 )$$}

\def\aligntag#1\endaligntag
  $${\gdef\tag##1\\{&(##1 )\cr}\eqalignno{#1\\}$$
  \gdef\tag##1$${\eqnono(\displayt@g##1 )$$}}

\def\eqalignno#1{\displ@y \tabskip\centering
  \halign to\displaywidth{\hfil$\displaystyle{##}$\tabskip\z@skip
    &$\displaystyle{{}##}$\hfil\tabskip\centering
    &\llap{$\displayt@gpar##$}\tabskip\z@skip\crcr
    #1\crcr}}

\def\displayt@gpar(#1){(\displayt@g#1 )}

\def\displayt@g#1 {\rm\ifunc@lled{#1}\global\advance\tagnumber by1
        {\def\next{#1}\ifx\next\empty\else\expandafter
        \xdef\csname @qnnum#1\endcsname{\t@ghead\number\tagnumber}\fi}%
  \writenew@qn{#1}\t@ghead\number\tagnumber\else
        {\edef\next{\t@ghead\number\tagnumber}%
        \expandafter\ifx\csname @qnnum#1\endcsname\next\else
        \w@rnwrite{Equation \noexpand\tag{#1} is a duplicate number.}\fi}%
  \csname @qnnum#1\endcsname\fi}

\def\ifunc@lled#1{\expandafter\ifx\csname @qnnum#1\endcsname\relax}

\let\@qnend=\end\gdef\end{\if@qnfile
\immediate\write16{Equation numbers written on []\jobname.EQN.}\fi\@qnend}

\catcode`@=12

\catcode`@=11
\newcount\r@fcount \r@fcount=0
\newcount\r@fcurr
\immediate\newwrite\reffile
\newif\ifr@ffile\r@ffilefalse
\def\w@rnwrite#1{\ifr@ffile\immediate\write\reffile{#1}\fi\message{#1}}

\def\writer@f#1>>{}
\def\referencefile{
  \r@ffiletrue\immediate\openout\reffile=\jobname.ref%
  \def\writer@f##1>>{\ifr@ffile\immediate\write\reffile%
    {\noexpand\refis{##1} = \csname r@fnum##1\endcsname = %
     \expandafter\expandafter\expandafter\strip@t\expandafter%
     \meaning\csname r@ftext\csname r@fnum##1\endcsname\endcsname}\fi}%
  \def\strip@t##1>>{}}

\def\citeall#1{\xdef#1##1{#1{\noexpand\cite{##1}}}}
\def\cite#1{\each@rg\citer@nge{#1}}     

\def\each@rg#1#2{{\let\thecsname=#1\expandafter\first@rg#2,\end,}}
\def\first@rg#1,{\thecsname{#1}\apply@rg}       
\def\apply@rg#1,{\ifx\end#1\let\next=\relax
\else,\thecsname{#1}\let\next=\apply@rg\fi\next}

\def\citer@nge#1{\citedor@nge#1-\end-}  
\def\citer@ngeat#1\end-{#1}
\def\citedor@nge#1-#2-{\ifx\end#2\r@featspace#1 
  \else\citel@@p{#1}{#2}\citer@ngeat\fi}        
\def\citel@@p#1#2{\ifnum#1>#2{\errmessage{Reference range #1-#2\space is bad.}
    \errhelp{If you cite a series of references by the notation M-N, then M and
    N must be integers, and N must be greater than or equal to M.}}\else%
 {\count0=#1\count1=#2\advance\count1
by1\relax\expandafter\r@fcite\the\count0,%
  \loop\advance\count0 by1\relax
    \ifnum\count0<\count1,\expandafter\r@fcite\the\count0,%
  \repeat}\fi}

\def\r@featspace#1#2 {\r@fcite#1#2,}    
\def\r@fcite#1,{\ifuncit@d{#1}          
    \expandafter\gdef\csname r@ftext\number\r@fcount\endcsname%
    {\message{Reference #1 to be supplied.}\writer@f#1>>#1 to be supplied.\par
     }\fi%
  \csname r@fnum#1\endcsname}

\def\ifuncit@d#1{\expandafter\ifx\csname r@fnum#1\endcsname\relax%
\global\advance\r@fcount by1%
\expandafter\xdef\csname r@fnum#1\endcsname{\number\r@fcount}}

\let\r@fis=\refis                       
\def\refis#1#2#3\par{\ifuncit@d{#1}
    \w@rnwrite{Reference #1=\number\r@fcount\space is not cited up to now.}\fi%
  \expandafter\gdef\csname r@ftext\csname r@fnum#1\endcsname\endcsname%
  {\writer@f#1>>#2#3\par}}

\def\r@ferr{\endreferences\errmessage{I was expecting to see
\noexpand\endreferences before now;  I have inserted it here.}}
\let\r@ferences=\references
\def\references{\r@ferences\def\endmode{\r@ferr\par\endgroup}}

\let\endr@ferences=\endreferences
\def\endreferences{\r@fcurr=0
  {\loop\ifnum\r@fcurr<\r@fcount
    \advance\r@fcurr by 1\relax\expandafter\r@fis\expandafter{\number\r@fcurr}%
    \csname r@ftext\number\r@fcurr\endcsname%
  \repeat}\gdef\r@ferr{}\endr@ferences}


\let\r@fend=\endpaper\gdef\endpaper{\ifr@ffile
\immediate\write16{Cross References written on []\jobname.REF.}\fi\r@fend}

\catcode`@=12

\citeall\refto          
\citeall\ref            %
\citeall\Ref            %

\catcode`@=11
\newwrite\tocfile\openout\tocfile=\jobname.toc
\newlinechar=`^^J
\write\tocfile{\string\input\space jnl^^J
  \string\pageno=-1\string\firstpageno=-1000\string\singlespace
  \string\null\string\vfill\string\centerline{TABLE OF CONTENTS}^^J
  \string\vskip 0.5 truein\string\rightline{\string\underbar{Page}}\smallskip}

\def\tocitem#1{
  \t@cskip{#1}\bigskip}
\def\tocitemitem#1{
  \t@cskip{\quad#1}\medskip}
\def\tocitemitemitem#1{
  \t@cskip{\qquad#1}\smallskip}
\def\tocitemall#1{
  \xdef#1##1{#1{##1}\noexpand\tocitem{##1}}}
\def\tocitemitemall#1{
  \xdef#1##1{#1{##1}\noexpand\tocitemitem{##1}}}
\def\tocitemitemitemall#1{
  \xdef#1##1{#1{##1}\noexpand\tocitemitemitem{##1}}}

\def\t@cskip#1#2{
  \write\tocfile{\string#2\string\line{^^J
  #1\string\leaderfill\space\number\folio}}}

%

\def\t@cproduce{
  \write\tocfile{\string\vfill\string\vfill\string\supereject\string\end}
  \closeout\tocfile
  \immediate\write16{Table of Contents written on []\jobname.TOC.}}


\let\t@cend=\endpaper\def\endpaper{\t@cproduce\t@cend}

\catcode`@=12

\tocitemall\head                

  \def\e{\epsilon} 
   
   \def\m{\mu}

\def \Dsl{\,\raise.15ex\hbox{$/$}\mkern-13.5mu {D}}
\def \DI{{\cal D}}
\def \dsl {\raise.15ex\hbox{$/$}\kern-.57em\hbox{$\partial$}}

\def \Del{\,\raise.15ex\hbox{$/$}\mkern-13.5mu {\Delta}}

\def \v {\rlap\slash{v}}

\def \el {\rlap\slash{\rm e}}

\ubec{92/15}
\title{Anomalies in the effective theory of heavy
quarks\footnote{$^*$}{\crgrant}}
\author{{\bf Joan Soto}\footnote\dag{soto@ebubecm1.bitnet} and
{\bf Rodanthy Tzani}\footnote\ddag{tzani@ebubecm1.bitnet}}
\affil{Departament d'Estructura i Constituents de la Mat\`eria,
Universitat de Barcelona, Barcelona, Catalonia, Spain}
\abstract{The question of the anomalies in the effective theory of
heavy quarks is investigated at two different levels. Firstly, it is
shown that none of the symmetries of this effective theory contains
an anomaly. The existence of a new `$ \gamma _ 5 $'-symmetry is pointed
out and shown to be also anomaly free. Secondly, it is shown
that the chiral anomaly of QCD is not reproduced in the effective
lagrangian for the heavy quarks, thus contradicting 't Hooft's anomaly
matching condition. Finally, the effective theory of heavy quarks is derived
from the QCD lagrangian in such a way that the terms leading to the anomaly
are included. For this derivation the generating functional method is used.}

\endtopmatter

\body
\baselineskip=20pt
{\bf I. Introduction}

The heavy quark effective theory (HQET), introduced originally by Isgur
and Wise [1], is a theory in which all the dependence on the large
rest mass
of the heavy quark has been removed analytically. In particular, one
assumes the heavy quark non-relativistic (essentially static) and
the theory is an expansion in the heavy
quark spacial momentum over its rest energy, that is an $ 1 \over m $
expansion [2].

The success of the formalism is based on the fact that the lowest order
in the expansion exhibits a number of new symmetries that are not
present in the original QCD lagrangian. One therefore can obtain model
independent results using only the symmetries of the theory in order to
relate matrix elements of physical processes [2,3].

There exist essentially two different approaches for the lagrangian
formulation of this effective theory. The first one, which is
conceptually the most direct, is based on the non-relativistic
approximation of the Dirac theory.
Performing directly a non-relativistic approximation in Dirac theory one
obtains the non-relativistic
Schr\"odinger-Pauli theory for a two component field, which describes
either
quarks or antiquarks, as an expansion
in $1 \over m $ [4].

The second formalism, due to Georgi [5], makes use of a new field $ h$
which contains only small fluctuations of the momentum. In other words,
the original quark field, redefined as
$$
\Psi = e^ { -i m \v v _{\mu} x ^ {\mu} }
h_ v (x) \,=\,( { 1 + \v \over  2 } ) e ^{-i m v \cdot x}
h ^+ _ v (x) + ( { 1 - \v \over 2}) e ^{ im v \cdot x} h ^- _ v(x)
\eqno(od)$$
is almost on
shell, except for small momenta fluctuations
described by the field $h_ v = h ^+_v + h ^-_v $. The
$4$-velocity of the system, $ v ^\m $,
is essentially the quark velocity in the $m \rightarrow \infty$ limit (
$v _\m v^\m = 1, v _ 0 <0$).
One then can show that the original QCD lagrangian
$$
i \bar \Psi \Dsl \Psi -m \bar \Psi \Psi
\eqno(o)$$
($D_{\mu}=\partial_{\mu}+B_{\mu}$ , $B_{\mu}$ being the $su(3)$-valued
gluon field) can be approximated
by an effective theory in terms of the field $h_{v}$, in which the mass
dependence has been removed from the Dirac operator. Namely,
$$
L _{eff} = i \bar h _ v \v v _\mu  D  ^ \mu h _ v
 = i \bar h_ v ^ +  \Dsl h _ v ^+ + i \bar h_v ^- \Dsl h _v ^- \,
\eqno(g)$$
where in the last expression $h_ v^+$ and $h _ v^- $ denote the quark
annihilation operator
and antiquark creation operator correspondingly and obey the relation $
\v h _ v ^{\pm} = { \pm } \; h  _v^ {\pm}$.
The subindex $v$ denotes the dependence of the quark fields on the
velocity. The basic idea of Georgi's effective
theory is to keep explicit the Lorentz covariance and hence exhibit all
the symmetries of the theory. This is, however, the lowest order in the
$1 \over m $ expansion and is not obvious how one should proceed
in order to include higher order corrections in the action.

In a later paper Mannel et al. [6] have derived
the effective lagrangian in the first order in $ 1 \over m $ using the
functional
integral approach. The advantage of this last formalism is that in
the rest frame it can be
shown to be equivalent with the non-relativistic approximation of
ref. [4], keeping, therefore, the physics more transparent.
In their formalism, however, it seems crucial to
keep in the theory only one kind of fields either quarks or
antiquarks, but not both.

In this work we shall restrict ourselves to purely theoretical aspects
of the heavy quark theory (see ref. [3] for a review on phenomenological
applications) which have not been discussed so far.
 In particular, we shall be concerned
with anomalies. We are going to address the issue at two distinct
levels. In the first one, we consider Georgi's effective action \(g) and
answer the question whether the flavor, spin, and an extra symmetry,
which we shall point out, are anomalous in the negative. This confirms
that the
symmetries of the effective theory are actually symmetries of the full
quantum theory so that they can be safely used to relate form factors as
it has been done so far. The second level addresses the question on how
the chiral anomaly of the fundamental theory is realized in the
effective theory.

This is a low energy
effective theory and according to 't Hooft consistency condition one
expects the same anomaly as in the original one.
On the other hand, since in this approximation the  heavy quark and
antiquark fields are separated by the infinite mass gap,
the original axial U(1) symmetry is lost. How then this low energy effective
theory realizes the chiral anomaly of the fundamental high energy theory?
In order to answer this
question we set a formalism based on the functional integral method.
In this way, the picture of the chiral anomaly becomes transparent.
We are also able to calculate the $ 1 \over m $ corrections of Georgi's
lagrangian in a conceptually simple and systematic way (i.e. we avoid
having to integrate out the antiparticles as in ref. [6]). Moreover, in
our
formalism it becomes clear that one can describe on the same footing any
number of flavors, quarks or antiquarks, with equal or different
velocities.

\bigskip\bigskip {\bf II. Anomalies in the symmetries of the
HQET}

In order to discuss the question of anomalies, we first analyze the
symmetries in Georgi's effective action \(g). In what follows we use
the notation of ref. [5].
There exist the following symmetries in this theory:
$$
h _ v \rightarrow e ^ {i \theta } \, h _ v\;\,\,\,{\rm and }\,\,\,\; h
 _ v \rightarrow e ^ {i \v \theta^{\prime}} \, h _ v \, ,
\eqno(s1)$$
with Noether's currents
$$
j ^\m \,=\, \bar h _ v  \v  v ^\m h  _ v \, \;\; {\rm and }\;\;\,
{j ^\m}^{\prime} \,=\, \bar h _ v  v ^\m h  _ v
\eqno(nc)$$
correspondingly.
These symmetries correspond to changing the field components $ h _ v ^+
$ and $ h _v ^-$ by arbitrary phases and result to the conservation of the
quark and antiquark number currents independently.
 \(s1) becomes the flavor symmetry
$$
{(h _v ^{\pm} )}^j \rightarrow {( e ^{i \theta _ {\pm}} )}^j _i
{(h_v ^{\pm} )}^i
\eqno(fs)$$
in a many flavor lagrangian.
Where now $ \theta _ {\pm} $ is an arbitrary traceless
hermitian matrix in the flavor space and in this case the currents (5)
are matrices in flavor space. There exist also the so called
`spin' symmetry in this lagrangian, as it is discussed in ref. [1] and
[5].
In particular, the action \(g) is invariant under the transformation
$$
h _v ^{\pm} \rightarrow \; e ^{ i \e ^i _{\pm}  S ^{\pm} _ i } \, h _
v ^{\pm} \;\; {\rm and} \;\;
\bar h_ v ^{\pm} \rightarrow \; \bar h _ v ^{\pm} \, e ^{- i \e_{\pm} ^i
S ^{\pm}_ i }
\eqno(ss)$$
where $ S^{\pm} _ i = i \e _ {ijk} [\el _j ,\el _k] (1 \pm \v) /2 $
and the parameters of the transformation $ \epsilon _ {\pm} ^i $ are
three arbitrary hermitian matrices in the flavor space. The orthonormal
set of the space like vectors $e ^\m _ j \, , \, j = 1, 2, 3 $ are
orthogonal to
$ v ^\m$. For one flavor, the classically conserved current due to the
spin symmetry is $$
j _i ^{\m, \pm} \,=\, \bar h ^{\pm} _ v \v v^\m S ^{\pm} _ i h^{\pm}
 _ v \,.
\eqno(sc)$$

For many flavors, a matrix in flavor space must be inserted in the
current (8). Notice that for the flavor and spin symmetries the fact
that we have both quark and antiquark fields in our theory is not
essential.
The above symmetries would persist even if we dropped either field.

It is not easy to keep track of the axial symmetry of the
original theory (2) in this approximation. The invariance
of \(o)
under the axial transformation manifests itself only in
the limit $ m \rightarrow 0 $.
It is broken explicitly by the mass term and in the limit $ m
\rightarrow \infty $, which is the limit of the effective theory, it can
not be realized in a simple way.

There exists, however, an extra symmetry in the HQET which formally
resembles the axial symmetry of the original theory being
nevertheless unrelated to it.
Classically the action of the HQET is invariant
under the transformation
$$
h _ v \rightarrow e ^ { i \gamma _ 5 \epsilon } h_v \,\,\,{\rm  and}
\;\; \bar h _ v \rightarrow \bar h _v \;\; e ^{ i \gamma _ 5 \epsilon }
\;,
\eqno(ns)$$
where again $\epsilon $ is a matrix in the flavor space.
It is worth noting that this symmetry is present only when one insists
in having the quark and antiquark fields with the same velocity.
For the rest of the discussion we shall refer to this symmetry as
`$\gamma _ 5$' in order to distinguish it from the axial symmetry of the
original theory.
Under this `$\gamma _ 5$'-symmetry the field components transform as
follows
$$
h _v ^{\pm} \rightarrow cos \epsilon \;\; h _v ^{\pm} + i \gamma _ 5
\; sin \epsilon  \;\; h _v ^{\mp}  \,.
\eqno(nsh)$$
This is an unexpected symmetry since it mixes quark and antiquark
fields, while in the lagrangian of the HQET the terms which mix quarks
and antiquarks
are neglected. This is a low energy effective theory which can not
describe heavy quark pair production, since it would require
very large momentum transfer.
The physical relevance of this last symmetry is not very
clear. The conserved quantity
corresponding to this symmetry has no obvious physical interpretation.
One can
show that the original axial symmetry coincides with the last $\gamma
_5$-symmetry in the limit $m \rightarrow 0$.
Indeed, under the axial transformation on the field $\Psi$,
the redefined field components transform as follows
$$
h _v ^{\pm} \rightarrow cos \epsilon \;\; h _v ^{\pm} + i \gamma _ 5 \;
sin \epsilon  \;\;e ^{ \pm 2im v \cdot x } h _v ^{\mp}  \,.
\eqno(de)$$
When $ m = 0 $ \(de) coincides with \(nsh), but
whenever $ m \ne 0 $ the two transformations become independent.
Since, however, \(od) is not a good field redefinition in the
limit $ m \rightarrow 0 $, in the sense that the lagrangian of the HQET
\(g)
cannot be derived in this limit, this coincidence can be superficial.

It is, now, easy to show that neither the spin nor the flavor symmetries
are anomalous in this theory. In both cases for the computation of the
divergence of the current, one has to compute the trace
of the commutator between the symmetry operator and the inverse propagator
$\Del$. Indeed
for the spin symmetry the relevant quantity is
$$
tr \left( \,[\, \e ^i _ {\pm} S ^{\pm} _ i \, ,\, \Del \,] \, {1 \over
\Del} \, G _ {reg} \, \right) \; ,
\eqno(sa)$$
where $\; \Del = \gamma ^ \mu \Delta _ \mu \;$ with $\;
\Delta _ \mu := v _ \mu v \cdot D \;$.
For the flavor symmetry it is enough to compute
$$
tr \left( \,[ \,\theta \,,\, \Del \,] \,{1 \over \Del} \,G _ {reg} \right)
 \;\;\;
{\rm and} \;\;\; tr \left( \,[\, \v {\theta} ^{\prime} \, ,\, \Del \,]
\,{1 \over \Del} \, G _ {reg} \,\right)
\eqno(fa)$$
where $\epsilon _ {\pm} ^{i}$ , $\theta $ and $\theta^{\prime}$ are
matrices in flavor space and
$\; G _ {reg} \;$ denotes the regulator.
Choosing then the regulator in such a way that it preserve the vector
symmetry, namely to be an appropriate function of $ \Del $, and using
trace properties the above commutators are zero.

For the case of the `$\gamma _
5$'-symmetry one needs to compute the analogous anticommutator,
in order to calculate the divergence of the corresponding Noether's
current
$$
j ^{\m} _ 5 \, = \, \bar h _ v\, \v \, v ^ {\m} \gamma _ 5 h _v \;.
\eqno(c)$$
(A matrix in flavor space must be inserted in \(c) for the case of
many flavors). In particular, the quantity to be computed is given by
$$
tr \left( \,\{ \, \epsilon (x) \gamma _ 5 , \Del \} \, {1 \over \Del }
G _{reg}\, \right) \,=\,
tr \left( \,2 \epsilon (x) \,\gamma _ 5 \,G_ {reg} \,\right)
\eqno(tr)$$
where again the regulator $G _{reg}$ is an appropriate function of $
\Del = \v v_{\mu} D ^{\mu}$.
Then, using trace properties, the fact that $ \v ^2 = 1$ and the
property
$\v \gamma _ 5 = - \gamma _ 5 \v $, it is easy to show that this gives
zero. Indeed, $$
\eqalign{
tr \left( \,2 \epsilon (x) \,\gamma _ 5 \,G_ {reg} \,\right) \,&=\,
tr \left( \,2 \epsilon (x) \, \v ^2 \, \gamma _ 5 \,G_ {reg} \,\right) \, = \,
 tr \left( \,2 \epsilon (x) \, \v \, \gamma _ 5 \,G_ {reg}\,\v \,\right) \, \cr
&=\,- tr \left( \,2 \epsilon (x) \, \v ^2 \, \gamma _ 5 \,G_ {reg} \,\right) \,
=\,0 \;. \cr }
\eqno(tr0)$$

With this we conclude that none of the symmetries in Georgi's action is
anomalous.

The physical reasoning behind the cancellation of the
`$\gamma _ 5$'- anomaly, is very interesting and it can be understood in
many different ways. In obtaining \(g) from \(o) one neglects in the
lagrangian terms that involve quark-antiquark pairs. Therefore, since
pair production is not relevant in this approximation, the
triangle graph is not possible in this effective theory and hence
there is no anomaly.

The phenomenon that the divergence of the $ j _ 5 ^ \mu $ current is
zero in this effective theory is related to taking the large
mass limit in the fermionic massive theory \(o). When $m$ becomes large,
the mass term in the action \(o) plays the role of
Pauli-Villars regulator with fields of opposite statistics. Therefore in
the limit $m \rightarrow \infty$ the contribution from the mass term in the
divergence of the axial current is equal in magnitude and
opposite in sign with the contribution from the anomaly. The two
contributions, therefore, cancel each other and the divergence
of the current is zero in this limit.
This could be a satisfactory explanation if the axial current
of \(o) coincided with the $ j _5 ^ \mu $ current of the effective
theory. In this case, however, the original axial current can not be related
to the $ j _ 5 ^ \mu$ in any straightforward way. On the other hand,
the effective theory of
heavy quarks is an expansion in $ 1 \over m$ and in principle
the matrix elements of physical processes could receive
contributions from the divergence of the current in some higher order
in $ 1 \over m$. The action \(g) is only a leading order in the
expansion and hence the dominant $1 \over m$ contributions in the
divergence of the current are missed in this approximation.
The question of the chiral anomaly, therefore, in this heavy quark
effective theory needs more investigation.
In what follows we analyze the chiral anomaly of the fundamental theory
in the heavy quark approximation.

\bigskip\bigskip {\bf III. The chiral anomaly in the HQET}

Once we have seen that none of the symmetries of the HQET is anomalous,
we now discuss the realization of the
$U_{L}(N_{f})\otimes U_{R}(N_{f})$ symmetry of the classical
QCD lagrangian with $N_{f}$ massless flavors in this effective theory.
This symmetry
is explicitly broken both by the quark masses and by the regularization.
The latter breaking leads to the chiral anomaly. Even
though the explicit breaking due to the quark masses is very large
for heavy quarks the question above is still relevant.
The chiral anomaly does not depend on the quark masses and breaks the
$U_{L}(N_{f})\otimes U_{R}(N_{f})$ symmetry down to the vector subgroup
$U_{V}(N_{f})$. (We understand that suitable sources are
added to the QCD lagrangian such that the above symmetries
become local, otherwise only the $U_{L}(1)\otimes U_{R}(1)$ part is
broken down to $U_{V} (1)$ by the anomaly whereas
$SU_{L}(N_{f})\otimes SU_{R}(N_{f})$ is spontaneously broken by
non-perturbative effects down to $SU_{V}(N_{f})$). Furthermore,
the chiral anomaly is given by one loop diagrams only and the anomalous
current conservation equation is not renormalized
at any order in perturbation theory [7]. These peculiar features
together with some important physical inputs related to the
inconsistency of chiral gauge theories [8] motivated the so-called
't Hooft anomaly matching condition [9] which in a wide sense (see [10]
for more precise discussions) can be stated as follows: any low-energy
effective theory must reproduce the chiral anomaly of the fundamental
theory from which it is a low energy realization. The 't Hooft
anomaly matching condition,
for instance, requires the introduction of a Wess-Zumino term in the
low energy realization of QCD with $N_{f}$ ($N_{f}> 2$) flavors by
chiral lagrangians [11]. The HQEL can be regarded as a low
energy realization of QCD for $m_{Q} >> \Lambda_{QCD}$. Adopting, then, the
point of view that 't Hooft anomaly matching conditions should be fulfilled
even for large mass theories, we investigate the question of chiral anomaly
in the HQEL.

Since $m_{Q}$ is large the explicit breaking of
$U_{L}(N_{f})\otimes U_{R}(N_{f})$ symmetry is also large in HQET.
In order to separate the explicit breaking due to $m_{Q}$ from
the breaking due to the regularization it is convenient to enlarge our
original theory by adding sources with appropriate transformation
properties under the gauge group in such a way that the action is explicitly
invariant [11]. Consider, then, our original lagrangian in Euclidean
space including the sources to be
$$
L=  \, \bar \Psi (\gamma^{\mu}D_{\mu}+\Phi )\Psi \, ,
\eqno(c1)$$
where $ D_{\mu}=  \partial_{\mu} + A_{\mu} \;\;\;\; $
and
$\; A_{\mu}:=  A_{\mu}^{L}P_{L}+A_{\mu}^{R}P_{R}\;$
is a $\; U _ L (N _f)\, \otimes \,U _ R (N _ f) \; $ gauge field (i. e.
$\, A_ {\mu} = A _
{\mu} ^ a T _ a \; $ with $ \, T _ a\,$ being the $\, U (N _f )\, $
generators).
$P _ {L, R} $ denote the projection operators,
$P_{L,R}=  {1\pm \gamma_{5} \over 2 } $.
In writing \(c1) we have promoted the mass term into a scalar field source
$\Phi$ defined by
$$
\Phi:=  \phi P_{L} -\phi^{\dagger}P_{R}  \;.
$$

Under the local chiral transformation, the fields $\, \Psi $, $ \bar
\Psi $ and $A _ {\mu} $ transform as
$$ \Psi  \longrightarrow g\Psi \,,\;\;\;
\bar \Psi  \longrightarrow \bar \Psi {\hat g}^{\dagger} \,,\;\;\;
A_{\mu}  \longrightarrow gA_{\mu}g^{\dagger} +
g\partial_{\mu}g^{\dagger}\,,
\eqno(c2)$$
where $g$ is defined by
$
g:= g_{L}P_{L}+g_{R}P_{R}\;\;\;\;{ \rm with} \;\;\;\;g_{L,R} \in U(N_{f})
$
and the meaning of hats is interchanging $P_{L}
\leftrightarrow P_{R}$ .

Then, the action \(c1) is invariant under \(c2) if we allow $ \Phi $ to
transform as
$$
\Phi  \longrightarrow \hat g \Phi g^{\dagger} \, .
\eqno (c21) $$
Our task is now to
generalize \(g) in such a way that (i) the chiral symmetry \(c2)  and \(c21)
is preserved
and (ii) \(g) is recovered upon switching off the sources, that is after
setting $A_{\mu}^{L}=A_{\mu}^{R}=B_{\mu}$ and $\Phi=iM$.
Since $\Phi^{\dagger}=-\hat \Phi$ we can write
$$
\Phi=i\hat U M U^{\dagger} \,  \quad\quad  {\rm with } \quad U^{\dagger}U=1 \,,
\eqno(c3) $$
where $M$ is a hermitian diagonal matrix and $U$ transforms as
$$
U \longrightarrow gU \;.
\eqno(c4)$$
Next, we define
$$
P_{\pm} := {1 \over 2} (1 \mp U\v U^{\dagger}) \, ,
\eqno(c5)$$
where $v_{\mu}$ is a diagonal matrix in the flavor space containing the
heavy quark velocities. It is easy to see that $ P _ {\pm} $ is a
projection operator, that is
$$
P_{\pm}^2=P_{\pm} \quad\quad ,\quad P_{+}P_{-}=P_{-}P_{+}=0
\quad\quad ,\quad \hat P_{\pm} \Phi = \Phi P_{\pm} \, .
\eqno(c6)$$

Then, the field redefinition \(od) is generalized to
$$
h_{v}^{\pm}  :=P_{\pm} U e^{ \mp i  v.x M} U^{\dagger} \Psi
\,\;\;\;{\rm and} \;\; \;
\bar h_{v}^{\pm}  :=\bar \Psi \hat P_{\pm} \hat U e^{ \pm i  v.x
M} \hat U^{\dagger} \eqno(c7)$$
and hence under the symmetry transformation \(c2) the quark fields
transform as
$$
h_{v}^{\pm}  \longrightarrow gh_{v}^{\pm} \;\;\; {\rm and} \;\;\;
\bar h_{v}^{\pm}  \longrightarrow \bar h_{v}^{\pm} {\hat
g}^{\dagger} \,.
\eqno(c8)$$
Substituting, now, \(c7) into \(c1) we obtain
$$
L \;=\; \bar h_{v}^{\pm} \; D_{v} ^ {\pm} \; h_{v}^{\pm}
\eqno(c9)$$
where
$$
\eqalign{D _ v ^ {\pm} \;&:=\; \hat P _ {\pm} \, [\,\pm \hat U v^{\mu}
U^{\dagger}\,(\,\partial_{\mu} \, +\, U\partial_{\mu}U^{\dagger}\,)\,
 \pm \,
 \hat U v^{\mu}\,e^{\mp iv.x M } U^{\dagger}D_{\mu}U e^{\pm i v.x M}
U^{\dagger}\, \cr
\,& +\, \gamma^{\mu} P_{\mp} U e^{\mp i v.x M } U^{\dagger}
D_{\mu} U e^{\pm iv.x M }U^{\dagger} ] P_ {\pm} \, . \cr }
\eqno(gl)$$
In obtaining the last expression we have used the relations
$$
\eqalign{& \hat P_{\pm} \gamma^{\mu} P_{\pm} \, = \, \pm \hat U v^{\mu}
U^{\dagger}P_{\pm}\, =\, \pm \hat P_{\pm}\hat U v^{\mu} U^{\dagger} \cr
&\partial_{\mu}\, (\,U \v U^{\dagger} \,) \, P_{\pm} \,= \, P_{\mp}
\partial_{\mu} \, (\,U \v U^{\dagger} \,)\, . \cr }
\eqno(c10)$$
The expression \(c9) with the definition \(gl) is a generalization of \(g)
and gives the
effective lagrangian for heavy quarks in a language in which the chiral
gauge invariance is explicit. Indeed, the action
is manifestly gauge invariant, since $U\partial_{\mu}
U^{\dagger} $ transforms as the gauge field $A_{\mu}$. Notice also
that \(c9) reduces to \(g) upon setting $U=1$ and $A_{\mu}=B_{\mu}$.

A key observation, now, is the following: Even though $ D_{v} ^ {\pm} $
transforms as the operator in
the original (anomalous) lagrangian \(c1), that is
$$
D_{v}^{\pm} \longrightarrow \hat g D_{v}^{\pm} g^{\dagger}
\eqno(c11)$$
\(c9) can be regulated in a
gauge invariant way due to the fact that now we have velocities at our
disposal. Indeed, due to the transformation property \(c11) the
eigenvalues of $ D _ v ^{\pm}$ are not chiral gauge invariant and,
therefore, appears that the determinant of the operator cannot be
regulated
in a chiral invariant way [12]. On the other hand, since $ \v ^2 = 1 $,
the following formal identities hold
$$
det(D ^{\pm}_{v})=1 \cdot det(D^{\pm}_{v})=det(\v^{\prime}) \cdot
det(D^{\pm}_{v})= det(\v^{\prime} D^{\pm}_{v})
\eqno(c12)$$
where $\v^{\prime}$ is a matrix proportional to the identity in flavour
space made out of one of the velocities in $\v $.
Now, under the transformation \(c2)
$$
{\v^{\prime}} D_{v} ^ {\pm} \longrightarrow g {\v^{\prime}} D_{v} ^
{\pm} g^{\dagger} \, .
\eqno(c13)$$
The last transformation property assures the chiral invariance of the
eigenvalues of the operator $ \v ^\prime D ^{\pm} _ v $ [12]. Therefore,
with an appropriate choice
of regulator, $ det ( \v ^\prime D^\pm _ v) $, and hence $ det ( D
^\pm _ v) $ because of \(c12), can be defined in a chiral invariant way.

The conclusion of the above analysis is that the lagrangian \(c9) of
the HQET does not reproduce the chiral anomaly of the
fundamental theory. This is related to the fact that the HQET is a
non-relativistic approximation (even if one writes it in a covariant
form) and it is crucial to insist in full relativistic covariance in
order to have chiral anomalies. If we
believe in 't Hooft anomaly matching
condition, then, the lagrangian \(c9) of the HQET must be modified in a
suitable way such that
the chiral anomaly is reproduced. We devote the rest of the paper to
this objective.

\bigskip\bigskip {\bf IV. A derivation of the HQET }

In this part we present an alternative derivation of the effective
theory of heavy quarks in such a way that we keep track of the chiral
anomaly of the fundamental high energy theory.
For this purpose we use the generating functional formalism.
Consider the generating functional of heavy quark Green
functions with insertions of heavy quark bilinear composite operators given by
$$
Z \, ( \bar \eta , \eta ; A _ {\Gamma} ) \, = \, \int D \bar \Psi D
\Psi \; exp \; [ i
\, \int _ x \, \left( \bar \Psi \;  {\DI}  \; \Psi \; + \; {\bar \eta}
\, \Psi \; + \; {\bar \Psi}  \eta \right) ]
$$
$$
=\, det ( \DI ) \; e ^{-i \int _ x \int _y {\bar \eta} (x) \, ( {\DI}
^{-1} ) _ {(x,y)} \, \eta (y) } \,\equiv \, e ^W \,,
\eqno(cg)$$
where we write explicitly only the piece of the
QCD lagrangian that contains heavy quark fields. Further integrations
over the gluon fields and the light quark fields with the rest of the
QCD lagrangian must be understood in \(cg) and in the formulas derived
from it.
The operator ${\DI}$ is given by
$$
{\DI} \,=\,i{\Dsl} \cdot  1 \, + \, \Gamma A _ \Gamma \, - \, M \, ,
\eqno(do)$$
where $ \Dsl \,=\, \gamma ^\m (  \partial _ \m + B _ \m )$, with
$B _ \m$ being the gluon field and $ \Gamma$ stands for any combination
of Dirac matrices. For definiteness
let us consider 2-flavor space (i.e., b and c quarks only), though the
analysis extends trivially to $N_{f}$ flavors.
The matrices $A _ {\Gamma}  $ and $M$ are given, then, by
$$
A _ {\Gamma}\,=\, \left(\matrix{ A ^{cc} _ \Gamma & A ^{cb} _ \Gamma \cr
A ^{bc} _ \Gamma & A ^{bb} _ \Gamma \cr } \right)
\quad\quad {\rm and} \quad\quad
M \,=\,\left(\matrix{  m _c & 0 \cr
0 &  m _ b \cr } \right) \,.
\eqno(mm)$$
$ A _ \Gamma $ can be either scalar,
axial, vector or axial vector source depending on the choice of $\Gamma$.
$ m _b $, $ m _c $ denote the quark masses and $ \bar \Psi $, $
\Psi $ and $ \bar \eta $, $\eta
$ are row and column matrices correspondingly in the space of flavors.

In order to describe heavy quarks almost on shell
we choose the following sources:
$$
\bar \eta (x) = {\bar \eta} ^+ _ v (x) \;\; {{1 + \v} \over 2} \; \;
e ^{i M v \cdot x}  \;\;{\rm and}\;\;
\eta (x) = e ^{-i M v \cdot x} \;\; { {1 + \v} \over 2} \;\; \eta ^+ _ v
(x) \,,
\eqno(so2)$$
where $\eta_{v}^{+}$ and $\bar \eta _ v ^ +$ are slowly varying functions
such that upon
functional variation will generate Green functions of heavy quarks with
small momentum fluctuations about $M \cdot v _ {\mu} $.
$v _ {\mu} $ is the following matrix of the quark velocities
$$
v _ {\mu} \,=\,\left(\matrix{ v _ \mu ^ c  & 0 \cr
0 & v _ \mu ^ b \cr } \right) \, ,
\eqno(vm)$$
where again $ v ^ 2 = 1$.
Note that with the choice of sources \(so2) we have chosen to describe only
quark fields in our theory. Extra antiquark or quark fields with different
velocities can be described by adding appropriate terms in our expressions
\(so2).

Next, we compute the generating functional $W$ except for the determinant.
The part of the generating functional to be computed is
$$
\int _ x \int _ y \, \bar \eta (x)  \; ( {\DI} ^{ -1} ) _ {(x,y)} \;
\eta (y) \,.
\eqno(rr)$$
Inserting \(so2) into \(rr) we obtain
$$
\int _ x \int _ y \, {\bar \eta} ^ + _ v (x)  \,
{ 1 + \v \over 2 } \, e ^ { i M v \cdot x } \,
({\DI} ^ {-1}) _ {(x,y)}\, e ^ { -i M v \cdot y} \, { 1 + \v \over 2 }
\, \eta _ v ^ + (y)
\eqno(lo)$$

We are interested in computing the inverse of the operator $ { \DI}$ as
a perturbation expansion in $ 1/M$.
We first notice that
$$
{ \DI } _ v \,:=\, e ^{ iM v.x} {\DI } e ^{ -i M v.x} \,=\,
i {\Dsl} \, - \, 2M P_- \,+\, \Gamma \alpha _ \Gamma  \, ,
\eqno(12)$$
where $ P_ {\pm}$ denote the following diagonal
matrices of the velocity projection operators
$$
P _{\pm} \, =\, \left(\matrix{ (1 \pm \v _ c ) /2 & 0 \cr
0 & ( 1 \pm \v _ b ) /2 \cr } \right)
\eqno(pm)$$
and $ \alpha _ \Gamma $ is a matrix whose elements are given by
$$
\alpha ^{ij} _ \Gamma \, =\, e ^{ i m _ i v _ i \cdot x } A ^{ij}
 _ \Gamma e ^{-i m _ j v_j \cdot x } \, .
\eqno(am)$$
$i,j$ take the values $c$ and $b$ correspondingly.
$\alpha^{ij}_{\Gamma}$
must be considered a slowly varying source as well. The reason for this
is that $A^{ij}_{\Gamma}$ must carry a momentum such that changes
$m_{j}v_{j}$ into $m_{i}v_{i}$ up to small momentum fluctuations [6].

In the expression \(12) the mass term contains zero modes, since it is
multiplied by the projection operator. In the presence of zero modes
the perturbation expansion around $ 1 \over M $ is not valid. In order
to avoid this problem, we first decompose the operator $ {\DI} _ v$
into the components $P _ \pm $ as follows
$$
{\DI } _ v \,=\,
\left(\matrix { i v \cdot D P _ + + P _+ \Gamma \alpha
_ \Gamma P _ + & P _ + (i \Dsl + \Gamma \alpha _ \Gamma ) P _ - \cr
P _- (i \Dsl + \Gamma \alpha _ \Gamma ) P _+ & (-i v \cdot D - 2M ) P _
- + P _ - \Gamma \alpha _ \Gamma P _- \cr } \right) \, ,
\eqno(sm)$$
where all the operators are matrices in the flavor space.
Again, at leading order,
when $ M \rightarrow \infty $, this operator is not invertible.
We must, therefore, first invert the operator and then perform the
$ 1 \over M $ expansion.

Next, we observe that the inverse ${ ({\DI}_v  )} ^{-1} $
can be written as
$$
{({\DI}_ v)} ^{-1} \,
=\, \left(\matrix { 1 & -R^ \prime \cr
0 & 1 \cr } \right)
\, \left(\matrix { A ^ {-1} & 0 \cr
0 & B ^ {-1} \cr } \right)
\, \left(\matrix { 1 & 0 \cr
-R  & 1 \cr } \right)
=\, \left(\matrix { A ^{-1} + R ^\prime B^{-1}  R &
-R^ \prime B ^{-1} \cr
- B ^{-1} R & B ^{-1} \cr } \right)
\eqno(ip)$$
where $A$, $B$, $R$ and $R ^ \prime $ are given by
$$
\eqalign{A \,&=\, i v \cdot D P_ + + P _ + \Gamma \alpha _ \Gamma P _+
\cr
B \,&=\, (-iv \cdot D - 2 M) P _- + P _- \Gamma \alpha _ \Gamma P _ - -
P _-(i \Dsl
+ \Gamma \alpha _\Gamma) {1 \over iv \cdot D + P _+ \Gamma \alpha
_\Gamma P _ + }
P_ + (i \Dsl + \Gamma \alpha _ \Gamma ) P _ - \cr
R \,&=\, P _- (i \Dsl + \Gamma \alpha _ \Gamma ) {1 \over iv \cdot D + P
_ + \Gamma \alpha _ \Gamma P _ + } P _ +  \cr
R ^\prime \,&=\, P _ + {1 \over i v \cdot D + P _+ \Gamma \alpha _
\Gamma P _ + } (i \Dsl + \Gamma \alpha _ \Gamma ) P _-   \cr }
$$
and the inverse, now, is defined in the space of projections, that is
$ A A ^{-1} = P _ +$, $B B ^ {-1} = P _ - $.

Then, we compute the inverse elements using perturbation theory and
obtain the following result for the effective action at subleading
order in $ 1 \over M $
$$
\int _ x \int _ y \bar \eta (x)  ( {\DI} ^{-1}) _ {(x,y)} \eta (y) \,=
\int _ x \int _ y {\bar \eta} ^+ _ v (x) \left\{ P _+ {1 \over { iv
\cdot D + P _+ \Gamma \alpha _ \Gamma P _ + }} P _ +  \right.
$$
$$
\left. - P _ + {1 \over i v \cdot D + P _ + \Gamma \alpha
_ \Gamma P _ + } (i \Dsl + \Gamma \alpha _ \Gamma) {1 \over 2 M }
P _ - (i \Dsl + \Gamma \alpha _ \Gamma ) {1 \over iv \cdot D + P _
+ \Gamma \alpha _ \Gamma  P _ + } P _ +  \right\} _ {(x,y)} \eta ^+ _v
(y) $$
$$
+ {\rm higher \; order \; terms.}
\eqno(effa)$$
The last expression gives the generating functional for heavy quarks
of velocity almost
$v$. In order to describe a system with additional quarks of different
velocity, for instance $v ^{\prime}$, one needs to add to $\eta _v$ in
\(so2) an analogous
component $\eta_{v^{\prime}}$. Then apart from the contribution \(effa)
and an analogous contribution for $v^{\prime}$ one gets terms mixing
$v$ and $v^{\prime}$. These terms consist of slowly varying
functions multiplying the
oscillating exponentials of the kind $exp{ \{-iM (v-v^{\prime}) \cdot x
\}}\;$.
We prove below that these terms can be put to zero at any finite order
in
$1/M$, as far as $M(1-v \cdot v^{\prime}) \gg \Lambda_{QCD} $.

The terms in question enter in the effective action as follows
$$
\int \, dx \, e ^ {-i M (v - v ^ \prime) \cdot x } f (x)
\eqno(ka)$$
where $f (x)$ is a slowly varying function. Equivalently \(ka) can
be written as
$$
\int _ x { 2 \over -i M (v - v ^ \prime ) ^ 2 }
( v _ \mu -v_{\mu}^{\prime}) \partial ^ \mu e ^ {-i M (v - v ^ \prime
) \cdot x}  f (x) \,=\, { 1 \over -i M (1 -v \cdot v ^ \prime ) }
\int _ x e ^ {-i M (v-v^\prime) \cdot x } (v _ \mu-v_{\mu}^{\prime})
\partial ^ \mu f(x) \,.\eqno(ot)$$
Since $f(x) $ varies slowly by iterating the process these terms can be
set to zero to any finite order in $ 1 \over M $.

Our expression for the generating functional \(effa) is non-local, since
it contains the propagator in the denominator. It can be shown, however,
that it can be derived from a local Quantum Field Theory.
Namely, the exponential of the right hand side of \(effa) can be written
as $$
{ 1 \over det ({\DI} _ v) }
\int  d \bar h ^ + _ v d h ^ + _ v
e ^ { i \int _ x \left(  \bar h ^ + _ v {\DI} _ v h ^ + _ v \,+
\, \bar \eta ^ +
_v h ^ + _ v \, + \, \bar h ^ + _ v \eta ^ + _ v \right) }
\eqno(last)$$
where $ {\DI} _ v $ is given by the following local expansion
$$
{\DI} _ v\, =\, i v \cdot D \,+\,P _ + \Gamma \alpha _ \Gamma P _ +
\,+\,
{1 \over 2 M } (i \Dsl + \Gamma \alpha _ \Gamma ) P _ - (i \Dsl + \Gamma
\alpha _ \Gamma ) \, + \,\;\; {\rm higher \; order \; terms}
\eqno(ce)$$
Then, using \(cg) and \(last) we obtain
$$
Z ( \bar \eta , \eta ; A _ \Gamma ) \,=\,{det {\DI} \over { det { \DI} _ v }}
\; \int  d \bar h ^ + _ v d h ^ + _ v
e ^ { i \int _ x \left( \bar h ^ + _ v {\DI} _ v h ^ + _ v \,+\,
\bar \eta ^ +
_v h ^ + _ v \, + \, \bar h ^ + _ v \eta ^ + _ v \right) } \, ,
\eqno(gfl)$$
with $ {\DI} _v $ given as in \(ce). In our last expression the
exponent at leading order coincides with the action \(g) and hence
$ h ^+ _ v $ can be identified with the heavy quark fields.

Let us ignore the determinants for a moment. By putting the sources
equal to the gluon field in the last exponent our local Quantum Field
Theory coincides with that given in ref.
[6]. In this respect our result is equivalent with the result obtained
by Mannel et al. as far as the above mentioned determinants can be
dropped. Furthermore, by taking functional derivatives of our expression
for the generating functional \(gfl) with respect to $ A ^ {ij}
_ \Gamma $ we generate the realization of all currents in the HQET.
Our results for the currents can be shown to
coincide with previous results at $O(1/M)$ [6], again if the
determinants can be dropped.

The determinants are the only ill-defined quantities in our generating
functional \(gfl). Keeping them explicit in our expressions
amounts to
keeping all the contributions leading to the chiral anomaly. In fact
$det(\DI)$ is known to give rise to the chiral anomaly [13]. On the
other hand
$det(\DI _v)$ can be defined in a chiral invariant way. This was shown in
the formulas \(c11) to \(c13) at leading order. In fact the proof holds
at any order
in $ 1/M$, since it is only based on the transformation properties of
$ {\DI}_ v$ and on the existence of the $ \v^{\prime}$ with $
{\v^{\prime}} ^ 2 = 1$.
This result is in agreement with ref. [6] where $ det({\DI} _ v) $ was
shown to be a constant by a different argument.
Therefore, the theory described by \(gfl) and \(ce) fulfills trivially the
't Hooft anomaly matching condition. The locality of the theory is
assured, since $det \DI$ admits a local expansion
in $1/M$ (we consider the Wess-Zumino term as local) the leading terms
of which have been given in ref. [13]. (See [14] for the case of
Majorana masses). Some next-to-leading
contributions have been recently calculated in [15]. We shall not
display them explicitly here.

The extra non-trivial contribution det${\DI}$, present in our approach,
has not played any role in the phenomenological
applications of the HQET considered so far, since its contribution to
the currents is
zero in the first order in $ 1/M$.
Nevertheless, it gives extra contributions to the currents in the HQET
which in principle can contribute in some higher order processes.

\bigskip\bigskip {\bf V. Conclusions}

In conclusion, we have examined the anomalies in the effective theory of
heavy quarks. We have shown that neither the spin nor the flavor
symmetries contain an anomaly in Georgi's lowest order
approximation of the effective theory for the
heavy quarks. Moreover, we have pointed out the existence of an extra
symmetry in this theory and have shown that this is also free of
anomaly.

In the second part of this work we investigated the question
of the chiral anomaly of the fundamental (QCD) theory. For this
purpose we have enlarged the original theory in such a way as to keep
the chiral symmetry explicit. We have, then, shown that the heavy
quark effective theory, obtained from this theory by just field
redefinition, is anomaly free.

Then, in order to keep track of any contribution due to the
anomaly, we have given an alternative derivation of the
effective theory of heavy quarks. In particular, using the
generating functional method we are able to compute in a sytematic
way higher order corrections in the $1/m$ expansion.
By explicitly keeping the determinant in this
path-integral
formalism, we account for all possible corrections due to the anomaly
in matrix elements.
The processes, however, in which the determinant would be relevant
typically correspond in QCD loop corrections to second
order electroweek processes and hence they are very suppressed. This
is probably the reason why our extra determinant contribution has been
unnoticed by previous authors.

Finally, we emphasize that our results obtained for the effective
currents and the local Quantum Field Theory coincide with previously
obtained results as far as the contributions from the determinant
can be ignored.
 \bigskip
 {\bf Acknowledgements}

We thank D. Espriu, A. M\"uller, A. P. Polychronakos, E. de Rafael and
A. Slavnov for discussions. We are specially indebted to J. Taron for
introducing us to the HQET, many discussions and a careful reading of
the manuscript.

                     \bigskip
\centerline{\bf References}

\item{[1]} N. Isgur and M.B. Wise, {\it Phys. Lett.} {\bf B232}
(1989) 113; and {\it Phys. Lett.} {\bf B237} (1990) 527.

\item{[2]} E. Eichen and B. Hill, {\it Phys. Lett.} {\bf B234} (1990)
511; M.E. Luke, {\it Phys. Lett.} {\bf B252} (1990) 447; A.F. Falk and
B. Grinstein, {\it Phys. Lett.} {\bf B247} (1990) 406; A.F. Falk and
B. Grinstein {\it Phys. Lett.} {\bf B249} (1990) 314;
A.F. Falk, B. Grinstein and M.E. Luke {\it Nucl. Phys.} {\bf B357}
(1991) 185; A. F. Falk, H. Georgi and B. Grinstein, {\it Nucl. Phys.}
{\bf B343} (1990) 1; H. Georgi {\it Nucl. Phys.} {\bf B348} (1991) 293;
H. Georgi, B. Grinstein and M.B. Wise {\it Phys. Lett.} {\bf B252}
(1990) 456; T. Mannel, W. Roberts and Z. Ryzak, {\it Phys. Lett.}
{\bf B271} (1991) 421 and {\it Phys. Rev.} {\bf D45} (1992) 875.

\item{[3]} B. Grinstein, preprint HUTP-91/A028 (1991) and
HUTP-91/A040, SSCL-Preprint-17 (1991).

\item{[4]} W.E. Caswell and G.P. Lepage {\it Phys. Lett.} {\bf B167}
(1986) 437;
G.P. Lepage and B.A. Thacker, {\it Nucl. Phys.} {\bf B4} (Proc.
Suppl.) (1988) 199.

\item{[5]} H. Georgi {\it Phys. Lett.} {\bf B240} (1990) 447.

\item{[6]}
T. Mannel, W. Roberts and Z. Ryzak, {\it Nucl. Phys.} {\bf B363},
(1991) 19.

\item{[7]} S.L. Adler and W.A. Bardeen, {\it Phys. Rev.} {\bf 182}
(1969) 1517.

\item{[8]} C. Bouchiat, J. Iliopoulos and Ph. Meyer,{\it Phys. Lett.}
{\bf B38} (1972) 519; D.J. Gross and R. Jackiw, {\it Phys. Rev.} {\bf
D6} (1972) 477.

\item{[9]} G. 't Hooft, in {\it ``Recent Developments in Gauge
Theories"}, G. 't Hooft et al. (Eds.), Plenum Press, New York, 1980.

\item{[10]} S. Coleman and B. Grossman, {\it Nucl. Phys.} {\bf B203}
(1982) 205.

\item{[11]} J. Gasser and H. Leutwyler, {\it Ann. Phys. } {\bf (N.Y.)
158} (1984) 142.

\item{[12]} L. Alvarez Gaum\'e and P. Ginsparg, {\it Nucl. Phys.} {\bf
B243} (1984) 449.

\item{[13]} R.D. Ball, {\it Phys. Rep.} {\bf 182} (1989) 1, and
ref. therein;
E. D'Hoker and E. Fahri, {\it Nucl. Phys.} {\bf B248} (1984)
77; R.D. Ball and H. Osborn, {\it Nucl. Phys.} {\bf B263} (1986) 245.

\item{[14]} J.L. Goity and J. Soto, {\it Phys. Lett.} {\bf B233} (1989)
400.

\item{[15]} J. Bijnens, {\it Nucl. Phys.} {\bf B367} (1991) 709.

 \vfill \end